\documentclass[a4paper,11pt]{article}
\usepackage[utf8]{inputenc}
\usepackage{color}
\usepackage{cite}
\usepackage{hyperref}
\usepackage{amsmath}
\usepackage{graphicx}
\usepackage[top=0.4in, bottom=0.7in, left=0.60in, right=0.5in]{geometry}

\title{{\bf\Large{Mass spectra and decay constants of heavy-light axial quarkonia in the framework of Bethe-Salpeter equation}}}
\author{Vaishali Guleria, Shashank Bhatnagar}
\begin{document}
\maketitle \small{Department of Physics, University Institute of Sciences, Chandigarh University, Mohali-140413, India}

\begin{abstract}
\normalsize{In this work we calculate the mass spectrum and decay constants of ground and excited states of heavy-light $P$-wave mesons such as $1^{++}$ and $1^{+-}$, with quark composition, $c\overline{u},
c\overline{s}, b\overline{u}, b\overline{s}$, and $b\overline{c}$
in the framework of a QCD motivated Bethe-Salpeter
equation (BSE) by making use of the exact treatment of the vector-vector type spin
structure in the
interaction kernel.  In this $4\times 4$ BSE framework, the coupled Salpeter equations for $Q\overline{q}$ are first
solved for the confining part of interaction, and are shown to decouple under
heavy-quark approximation. Then the
one-gluon-exchange interaction is perturbatively incorporated,
leading to their mass spectral equations. The analytic forms of
wave functions obtained from these equations are then used for
calculation of leptonic decay constants of ground and excited
states of $1^{++}$, and $1^{+-}$ as a test of these wave functions
and the dynamical framework.}
\end{abstract}
\bigskip
Key words: Bethe-Salpeter equation, Salpeter equations, Mass
spectral equation, Heavy-Light Quarkonia, Decay constants

\large{
\section{Introduction}
 There is a growing interest in the
experimental and theoretical studies of heavy-light mesons over the last few years. Studies on heavy-light mesons are important
for the determination of Cabibo-Kobayashi-Maskawa (CKM) mass
matrix elements. These studies on quarkonia need heavy quark
dynamics, which can provide a significant test of Quantum
Chromodynamics (QCD). A number of non-perturbative approaches, such as
NRQCD\cite{brambilla05}, QCD sum rule\cite{shifman79}, potential
models\cite{quigg19, Godfrey85,Ebert03}, lattice QCD\cite{colquhoun15,Bali97,Burch10},
Bethe-Salpeter equation (BSE)
method\cite{Mitra01,Mitra92,Munczek93,bhatnagar14,bhatnagar06,bhatnagar11,glwang},
heavy quark effective theory\cite{neubert94}, Relativistic
Quantum Model (RQM)\cite{ebert11}, and Chiral perturbation
theory\cite{gasser84}  have been employed to study spectroscopy of heavy quarkonia.

A lot of investigation has been carried out on S-wave mesons such as pseudoscalar ($0^{-+}$), and vector ($1^{--}$) mesons.
While comparatively lesser investigation has been carried out on $P$-wave mesons such as
scalar ($0^{++}$), and axial (both $1^{++}$ and $1^{+-}$) mesons. Further, new states are continuously being discovered at experimental facilities around the world. Some of the recently discovered $P$-wave conventional quarkonium states are: $h_b(1P), h_b(2P), \chi_{b1}(3P)$ and $\chi_{b2}(3P)$\cite{olive14,patrignani16}. The discoveries of conventional states $h_c(1P), h_c(2P),
\chi_c(1P), \chi_c(2P)$, and $\eta_c(1S)$, and the observation of the exotic states like $X(3872), X(3915), Y(4260), Z(3930)$ at
Belle, BaBar, LHC, BESIII,CLEO, etc have created a renewed interest in quarkonium physics \cite{olive14}. These exotic states are the unconventional states that were named($X,Y,Z$), that do not fit into conventional quarkonium states. Amongst these states are $Z_c(3900)$ and $Z_c(4430)$, whose quantum number assignments have been confirmed to be $1^{+-}$ as per the Particle Data Group (PDG) 2020 tables\cite{zyla2020}. Further there is a lack of knowledge of decay constants of axial meson states in general.

Thus, this work is devoted to calculation of mass spectrum and leptonic decay constants of ground and excited states of $P$-wave heavy-light ($Q\bar{q}$) mesons, such as axial vector quarkonia, which fall into two classes: $1^{++}$, and $1^{+-}$. Here, $1^{+-}$ mesons have $L=1$, but no spin excitation (i.e. $S=0$), while $1^{++}$ have $L=1$, and $S=1$. This give total angular momentum, $J=1$ for both types of mesons. This difference gives these two classes of mesons slightly different masses.

The $1^{+-}$ state was first detected in $p\bar{p}$ collisions by $R704$ collaboration \cite{r704}. $h_c$ was then found in reaction, $p\bar{p}\rightarrow h_c\rightarrow\pi^0 J/\Psi\rightarrow e^+ e^-$ in FNAL E760 experiment, with mass, $M_{h_c}=3526.2\pm 0.2\pm 0.2 MeV.$, and $\Gamma_{h_c}< 1.1 MeV.$\cite{FNAL}. Very  recently, BES III collaboration reported $h_c$ production in the process,  $e^+ e^- \rightarrow \pi^+ \pi^- h_c$\cite{BESIII}.

$1^{++}$ mesons are seen in $pp$ collisions. However, not many decays of these mesons are experimentally observed as can be checked from PDG tables \cite{zyla2020}. In the present work we thus study mass spectrum, and the leptonic decays of ground and excited states of $1^{+-}$ and $1^{++}$ $Q\bar{Q}$, and $Q\bar{q}$ quarkonia. Though the mass spectrum of these quarkonia have been experimentally observed, but so far there is no experimental determination of their leptonic decays\cite{zyla2020} so far, though calculations of these quantities have been studied in some models. Our predictions of leptonic decay constants of these mesons may provide a guide to future experiments for determination of these values. Further, the values of leptonic decay constants that we calculate in this work play might play an important role in precise determination of CKM matrix elements, studies of weak decays and meson mixing.

We further wish to study the single photon radiative decays involving these axial mesons through processes such as, $A^+\rightarrow V\gamma$, $A^-\rightarrow P\gamma$, $V\rightarrow A^+\gamma$, and $P\rightarrow A^-\gamma$ ($A^+/ A^-= 1^{++}/1^{+-}$ (axial vector), $V=1^{--}$ (vector), $P=0^{-+}$ (pseudoscalar), and $S=0^{++}$ (scalar) quarkonia), which have been studied by some models\cite{deng17}, for which experimental data\cite{olive14} is available, having recently studied the radiative $M1$ and $E1$ transitions, such as $V\rightarrow P\gamma$; $V\rightarrow S\gamma$, and $S\rightarrow V\gamma$ \cite{bhatnagar20}. The transitions involving leptonic and radiative decays of axial vector quarkonia would also serve as a test for the analytic forms of wave functions of these mesons calculated analytically in this paper by solving their mass spectral equations.

Here, regarding the mass spectral calculations, we wish to mention that in unequal mass systems
such as $Q\overline{q}$, the quarks are not very close together. Due to this
the confining interaction would dominate over the One-Gluon-Exchange (OGE) interactions for $Q\overline{q}$ systems.
Thus, the perturbative incorporation of OGE term \cite{bhatnagar18,gebrehana19} is a reasonable
approximation for heavy-light quarkonia.

To study heavy-light quarkonia, we make use of unequal mass kinematics, where the Wightman-Garding definitions of internal momenta of quarks act as the momentum partition parameters. We make use of the four effectively 3D Salpeter equations (that are obtained through the 3D reduction of the 4D Bethe-Salpeter equation under Covariant Instantaneous Ansatz, which is a Lorentz-invariant generalization of Instantaneous Approximation. We use the full Dirac structure for writing down the wave functions of these hadrons in accordance with \cite{smith, alkofer01}.
We first solve the Salpeter equations and obtain coupled equations in amplitudes of Dirac structures, which are then decoupled under heavy-quark approximation, and are used to obtain mass spectrum that is explicitly dependent on the principal quantum number, $N$ in an approximate harmonic oscillator basis.  We then incorporate perturbatively the one-gluon-exchange interaction, and obtain not only the mass spectrum, but also the algebraic forms of wave functions in approximate harmonic oscillator basis. These wave functions are then used for analytic calculations of leptonic decay constants of both $1^{+-}$ and $1^{++}$ mesons. These analytic calculations have advantage of transparency in expressing spectrum in terms of principal quantum number, $N$, and also obtaining algebraic forms of wave functions that can be employed for calculations of various transitions involving these mesons. Thus, we rely on purely analytic approach in contrast to most approaches in the literature that resort to numerical analysis at an early stage.

This paper is organized as follows: In section 2, we introduce the
formulation of the $4\times 4$ Bethe-Salpeter equation under the
covariant instantaneous ansatz, and derive the hadron-quark
vertex. In sections 3 and 4,  we derive the mass spectral
equation of heavy-light $1^{+-}$, and $1^{++}$ mesons
respectively. In section 4 is devoted to the calculations of their decay constants, Section 5 is devoted to numerical
results and discussion.

\section{Formulation of the 4$\times$ 4 Bethe-Salpeter equation}

Our work is based on QCD motivated Bethe-Salpeter equation in ladder approximation. It makes use of an effective four-fermion interaction that is mediated
by a gluonic propagator, which is employed as the kernel of BSE in the lowest order. Our interaction kernel comprises of a confining term and a one-gluon exchange term. The effective forms of the BS kernel in ladder approximation have recently been used in
\cite{karmanov19,glwang,fredrico14,hea19,wang19}. They can be used to study relativistic bound states. As mentioned above, the frame work of Bethe-Salpeter equation is quite general, and provides an effective description of bound quark- antiquark systems through a
suitable choice of input kernel for confinement.

We summarize the main points about the 4$\times$ 4 Bethe-Salpeter equation under the
Covariant Instantaneous Ansatz (CIA). Here, CIA which is a Lorentz-invariant
generalization of Instantaneous Approximation (IA), which is used
to derive the 3D Salpeter equations\cite{yang,hluf16,bhatnagar18,gebrehana19,bhatnagar20}.
We start with a 4D BSE for quark- anti quark system with quarks of
constituent masses, $m_{1}$ and $m_{2}$, written in a $4\times 4$
representation of 4D BS wave function $\Psi(P,q)$ as:

The Bethe-Salpeter equation that describes the quark-anti quark bound state of momenta $p_1$ and $p_2$, relative
momentum $q$, and meson momentum $P$ is
\begin{equation}
S_{F}^{-1}(p_{1})\Psi(P,q)S_{F}^{-1}(-p_{2}) =
i\int \frac{d^{4}q'}{(2\pi)^{4}}K(q,q')\Psi(P,q'),
\end{equation}
where $K(q,q')$ is the interaction kernel, and $S_{F}^{-1}(\pm p_{1,2})=\pm i{\not}p_{1,2}+ m_{1,2}$ are the quark/antiquark propagators, where the momenta of quarks can be expressed as,

\begin{equation}
p_{1,2}=\hat{m}_{1,2}P \pm q,
\end{equation}

where $\hat{m}_{1,2}=1/2[1\pm (m_1^2-m_2^2)/M^2]$ act as momentum partitioning parameters. We now make use of the Covariant Instantaneous Ansatz, where,
$K(q,q')=K(\widehat{q},\widehat{q}')$ on the BS kernel, where the BS kernel depends entirely on the variable,
$\widehat{q}_\mu= q_\mu- \frac{q.P}{P^2}P_\mu$. Here, $\widehat{q}$ the transverse component of
internal momentum of the hadron, that is orthogonal to the total
hadron momentum ($\widehat{q}.P=0$), and $\sigma
P_\mu=\frac{q.P}{P^2}P_\mu$ is the longitudinal component of $q$, that is parallel
to $P$. Here, the 4-dimensional volume element is,
$d^4q=d^3\widehat{q}Md\sigma$. Now working on the right side of Eq.(1), and making use of the fact that

\begin{equation}
\psi(\hat{q}')=\frac{i}{2\pi}\int Md\sigma' \Psi(P,q'),
\end{equation}

and the fact that the longitudinal component of $Md\sigma$  of $q$ does not appear in $K(\hat{q}, \hat{q}')$, carrying out integration over $Md\sigma$ on right side of Eq.(1), we obtain,

\begin{equation}
S_{F}^{-1}(p_{1})\Psi(P,q)S_{F}^{-1}(-p_{2})=\int \frac{d^3\hat{q}'}{(2\pi)^3}K(\hat{q}, \hat{q}')\psi(\hat{q}'),
\end{equation}

We can express the 4D BS wave function, $\Psi(P,\hat q)$ as,
\begin{equation}
 \Psi(P,\hat q)=S_1(p_1)\Gamma(\hat q)S_2(-p_2),
\end{equation}

Here the 4D hadron-quark vertex, that enters into the definition of the 4D BS wave function in the previous equation, that be identified as,
\begin{equation}\label{6a}
 \Gamma(\hat q)=\int\frac{d^3\hat q'}{(2\pi)^3}K(\hat q,\hat q')\psi(\hat q').
\end{equation}
Thus, it can be observed that $\Gamma(\hat{q})$, as seen from the Eqs.(4), (5), is directly related to the 4D wave function, $\Psi(P,q)$, and one can express the 4D BS wave function $\Psi(P,q)$ in terms of $\Gamma(\hat{q})$.

Following a sequence of steps outlined in \cite{hluf16}, we get four Salpeter equations,

\begin{eqnarray}
 &&\nonumber(M-\omega_1-\omega_2)\psi^{++}(\hat{q})=\Lambda_{1}^{+}(\hat{q})\Gamma(\hat{q})\Lambda_{2}^{+}(\hat{q})\\&&
   \nonumber(M+\omega_1+\omega_2)\psi^{--}(\hat{q})=-\Lambda_{1}^{-}(\hat{q})\Gamma(\hat{q})\Lambda_{2}^{-}(\hat{q})\\&&
\nonumber \psi^{+-}(\hat{q})=0.\\&&
 \psi^{-+}(\hat{q})=0\label{fw5}
\end{eqnarray}

Now, from the first two of the Salpeter equations above that are used for the mass spectral calculations, one can see the appearance of the hadron-quark vertex function, $\Gamma(\hat{q})$ that is employed for the transition amplitude calculations in 4D basis on their right hand sides. This gives an interconnection between the 3D wave function, $\phi(\hat{q})$ that satisfies the Salpeter equations and the 4D hadron-quark vertex $\Gamma(\hat{q})$, This provides an important link between the (low energy) spectra and (high energy) transition amplitudes, which is an important feature about this BSE framework.

The BSE interaction kernel with spin, colour and orbital parts is written as,
\begin{equation}\label{cf0}
 K(\hat q',\hat q)=(\frac{1}{2}\vec\lambda_1.\frac{1}{2}\vec\lambda_2)(\gamma_\mu\otimes\gamma_\mu)V(\hat q',\hat q).
\end{equation}
The hadron-quark vertex function can be expressed as \cite{bhatnagar18},
\begin{equation}\label{6aa}
 \Gamma(\hat q)=\int\frac{d^3\hat q'}{(2\pi)^3}V(\hat q,\hat q')\gamma_\mu\psi(\hat q')\gamma_\mu.
\end{equation}

Here, the 3D BS wave function,
$\psi(\widehat{q})$ is flanked by two  $\gamma_{\mu}$s. And the scalar part of the kernel,
$V=V_{OGE}+ V_{Confinement}$, whose expression is,

\begin{eqnarray}\label{fr1}
&&\nonumber V(\hat q,\hat q')=\frac{4\pi\alpha_s}{(\hat q-\hat
q')^2}
 +\frac{3}{4}\omega^2_{q\bar q}\int d^3r\bigg(\kappa r^2-\frac{C_0}{\omega_0^2}\bigg)e^{i(\hat q-\hat q').\vec r},\\&&
 \kappa=(1+4\hat m_1\hat m_2A_0M^2r^2)^{-\frac{1}{2}}.
\end{eqnarray}

The QCD motivation to the kernel is provided by the dependence of the flavour dependent spring constant $\omega_{q\bar{q}}^2$ on the running QCD coupling constant, $\alpha_s$. The smooth transition from nearly harmonic (for $c\bar{u}$) to almost linear (for $b\bar{b}$)} is guaranteed by the algebraic form of the confining potential employed in this work.

The structure of the confinement part $V_{c}(\hat q,\hat q')$ in terms of $\bar{V}_c$, and its structure of is taken from \cite{hluf16,gebrehana19}.
The framework is quite general so far. To evaluate the mass spectral equations, we have to use the four Salpeter equations in Eqs.(7).\\

\section{Mass spectral equation for heavy-light $1^{+-}$ quarkonia}

We start with the general form of 4D BS wave function for axial
meson ($1^{+-}$) in \cite{smith,alkofer01}. Taking its dot product with $\epsilon_{\mu}$, the polarization vector of axial vector meson, we get,

\begin{equation}
\Psi_{A^-}(P,q)=\gamma_5(q.\epsilon)[g_1(q,P)+i\not{P} g_2(q,P)-i\not{q} g_3(q,P)+[\not{P},\not{q}]g_4(q,P)].
\end{equation}

Then, making use of the 3D
reduction and making use of the fact that $\widehat{q}.P=0$, we
can write the general decomposition of the instantaneous BS wave
function for scalar mesons $(J^{pc}=1^{+-})$, of dimensionality
$M^0$ being composed of various Dirac structures that are multiplied
with scalar functions $g_i(\hat q)$ and various powers of the
meson mass $M$ as \cite{bhatnagar18}

\begin{equation}\label{uw1}
 \psi_{A^-}(\hat q)=\gamma_5 \frac{(\epsilon.\hat{q})}{M}[g_1(\hat{q})+i\frac{{\not}P}{M}g_2(\hat{q})-i\frac{{\not} \hat{q}}{M}g_3(\hat{q})+2\frac{{\not} P{\not} \hat{q}}{M^2}g_4(\hat{q})].
 \end{equation}

Till now these amplitudes $g_{1},...,g_{4}$ in equation above
are all independent, and as per the power counting rule
\cite{bhatnagar06,bhatnagar14} proposed by us earlier, the $g_1$,
and $g_2$ are the amplitudes associated with the leading Dirac
structures, namely $\gamma_5I$ and $\gamma_5{\not}P/M$, while $f_3$ and $f_4$ will
be the amplitudes associated with the sub-leading Dirac
structures, namely, $\gamma_5{\not}\hat{q}/M$, and
$\gamma_5\frac{2{\not}P{\not}\hat{q}}{M^2}$.

We now use the last two Salpeter equations $\psi^{+-}(\hat q)=
\psi^{-+}(\hat q)=0$ in Eq.(\ref{fw5}), that can be used to obtain the
constraint relations between the scalar functions for unequal mass
mesons. We wish to mention that due to the two constraint equations, the
scalar amplitudes, $g_i(\widehat{q}) (i= 1,...,4)$ are no longer all
independent, but are tied together by the relations.

\begin{eqnarray}
&&\nonumber g_4 =\frac{M(m_1\omega_2-m_2\omega_1)}{2(\omega_1+\omega_2)\hat{q}^2}g_2,\\&&
g_3=\frac{(\hat{q}^2+\omega_1\omega_2-m_1m_2)M}{\hat{q}^2(m_1+m_2)}g_1.
\end{eqnarray}

Making use of the above relations between amplitudes, we can write the complete 3D Salpeter wave function, $\psi_{A^-}(\hat{q})$ as,

\begin{equation}
\psi_{A^-}(\hat{q})=\gamma_5 \frac{\epsilon.\hat{q}}{M}\bigg[g_1\bigg(1- i\not{\hat{q}\frac{(\hat{q}^2+\omega_1 \omega_2-m_1 m_2)}{\hat{q}^2 (m_1+m_2)}\bigg)+g_2\bigg(i\frac{\not{P}}{M}+\frac{(m_1\omega_2-m_2\omega_1)}{(\omega_1+\omega_2)\hat{q}^2}\frac{{\not}P{\not}\hat{q}}{M}}\bigg)\bigg].
\end{equation}

We proceed in the same way as, \cite{bhatnagar18}, where on the right side of these equations, we
first work with the confining interaction, $V_c(\widehat{q})$ alone.
The coupled integral equations that result from the first two Salpeter equations are:

\begin{eqnarray}
&&\nonumber (M-\omega_1-\omega_2)[2g_1 +g_2 L]= \int \frac{d^3\hat{q}'}{(2\pi)^3}V_c(\hat{q}')[g_1 H'_1 +g_2 H'_2]\\&&
\nonumber (M+\omega_1+\omega_2)[2g_1 -g_2 L] = \int \frac{d^3\hat{q}'}{(2\pi)^3}V_c(\hat{q}')[g_1 H'_1 -g_2 H'_2]\\&&
\nonumber L=\frac{m_1}{\omega_1}+\frac{m_2}{\omega_2}+\frac{(m_1\omega_2-m_2\omega_1)}{\omega_1+\omega_2}(\frac{1}{\omega_2}-\frac{1}{\omega_2})\\&&
\nonumber H'_1=-4-4\frac{m_1m_2+\hat{q}'^2}{\omega'_1\omega'_2}+2\frac{(m_1-m_2)}{m_1+m_2}(\hat{q}'^2+\omega'_1\omega'_2-m_1m_2)\\&&
H'_2=2(\frac{m_1}{\omega'_1}-\frac{m_2}{\omega'_2}),
\end{eqnarray}

where, $\omega'^2_{1,2}=m^2_{1,2}+\hat{q}'^2$, and $H'_{1,2}$ involve $\hat{q}'$. Now, making use of the structure of confining interaction, $\bar{V}_c(\hat{q}')=\bar{V}_c(\hat{q}\delta^3(\hat{q}-\hat{q}')$ in the model, we can express the above two coupled integral equations into two coupled algebraic equations,

\begin{eqnarray}
&&\nonumber (M-\omega_1-\omega_2)[2g_1 +g_2 L]= V_c(\hat{q})[g_1 H_1 +g_2 H_2]\\&&
(M+\omega_1+\omega_2)[2g_1 -g_2 L] = V_c(\hat{q})[g_1 H_1 -g_2 H_2],
\end{eqnarray}

where $H_{1,2}$ now involve $\hat{q}$. To decouple the above algebraic equations, we first add these equations, and substract the second equation from the first. We thus get two algebraic equations that are again coupled in $g_1$, and $g_2$. Eliminating $g_1$ in terms of $g_2$ from the first equation, and putting in the second equation, and similarly, eliminating $g_2$ in terms of $g_1$ from the second equation, and putting in the first equation, we get two identical decoupled equations in $g_1$, and $g_2$ as:

\begin{equation}
\begin{split}
  \bigg[\frac{M^2}{4}-\frac{1}{4}(m_1+m_2)^2-\hat q^2\bigg]g_1(\hat q)&=-\frac{1}{2}(m_1+m_2)\overline{ V}_{c}(\hat q)g_1(\hat q)\\
 \bigg[\frac{M^2}{4}-\frac{1}{4}(m_1+m_2)^2-\hat q^2\bigg]g_2(\hat q)&=-\frac{1}{2}(m_1+m_2)\overline{ V}_{c}(\hat q)g_2(\hat q).
\end{split}
\end{equation}

Since the two equations are of
the same form in scalar functions $g_1(\hat q)$ and $g_2(\hat q)$,
that are the solutions of identical equations, we can take,
$g_1(\hat q)\approx g_2(\hat q)(=\phi_A(\hat q))$. Thus, we can write the complete wave function for $1^{+-}$ meson as,

\begin{equation}
\psi_{A^-}(\hat{q})=\gamma_5 \frac{\epsilon.\hat{q}}{M}\bigg[\bigg(1- i\not{\hat{q}\frac{(\hat{q}^2+\omega_1 \omega_2-m_1 m_2)}{\hat{q}^2 (m_1+m_2)}\bigg)+\bigg(i\frac{\not{P}}{M}+\frac{(m_1\omega_2-m_2\omega_1)}{(\omega_1+\omega_2)\hat{q}^2}\frac{{\not}P{\not}\hat{q}}{M}}\bigg)\bigg]\phi_{A^-}(\hat{q}).
\end{equation}

Using the expression for $\overline{V}_c(\hat q)$ given above, we get the
equation,

\begin{equation}\label{s23}
 E_A\phi_{A^-}(\hat q)=[-\beta_{A^-}^4 \vec \nabla^2_{\hat q} +\hat q^2]\phi_{A^-}(\hat q),
\end{equation}

where the inverse range parameter $\beta_A^-$ can be expressed as,

\begin{equation}\label{be}
\begin{split}
\beta_{A^-}&=(\frac{2}{3}(m_1+m_2)\omega_{q\bar{q}}^2)^{\frac{1}{4}},\\
\omega_{q\bar q}&=(4M\hat m_1\hat m_2\omega_0^2\alpha_s(M))^{1/2},\\
\alpha_s&=\frac{12\pi}{33-2N_f}\log\bigg(\frac{M^2}{\Lambda_{QCD}^2}\bigg)^{-1}
\end{split}
\end{equation}

The solutions of Eq.(\ref{s23}) are calculated by using the power series method. We assume the solutions of Eq.(19) are of the form, $\phi(\hat{q})=\xi(\hat{q})e^{-\frac{\hat{q}^2}{2\beta_{A^-}^2}}$. Then Eq.(19) can be expressed as,

\begin{equation}
\xi''(\hat{q})+(\frac{2}{\hat{q}}-\frac{2\hat{q}}{\beta_{A^-}^2})\xi'(\hat{q})+(\frac{E}{\beta_{A^-}^4}-\frac{3}{\beta_{A^-}^2}-\frac{l(l+1)}{\hat{q}^2})\xi(\hat{q})=0.
\end{equation}

The energy eigen values of this equation obtained using the power series method are:

\begin{eqnarray}
&&\nonumber E_N= 2\beta_{A^-}^2 (N+\frac{3}{2});\\&&
N=2n+l
\end{eqnarray}

with the quantum number $n$ taking values, $n=0,1,2,...$, and the orbital quantum number $l=1$ that corresponds to $P$ wave states. This leads to the mass spectral equation for axial vector ($1^{+-})$ quarkonia as,

\begin{equation}\label{mse0}
\frac{1}{4}\bigg[M^2-(m_1+m_2)^2\bigg]+\frac{C_0\beta_{A^-}^4}{\omega_0^2}\sqrt{1+8\hat
m_1\hat
m_2A_0(N+\frac{3}{2})}=2\beta_{A^-}^{2}(N+\frac{3}{2}),~~~N=1,3,5,...,
\end{equation}

It can further be checked that for each value of $n=0,1,2,...$, would thus correspond a polynomial, $\xi(\hat{q})$ of order $2n+1$ in $\hat{q}$. These are obtained as solutions of Eq.(22). The normalized odd-parity eigen functions derived as solutions of Eq.(19) are:

\begin{equation}\label{wv1}
\begin{split}
 \phi_{A^-}(1P,\hat q)&=\sqrt{\frac{2}{3}}\frac{1}{\pi^{3/4}}\frac{1}{\beta_{A^-}^{5/2}} \hat qe^{-\frac{\hat q^2}{2\beta_{A^-}^2}}\\
 \phi_{A^-}(2P,\hat q)&=\sqrt{\frac{5}{3}}\frac{1}{\pi^{3/4}}\frac{1}{\beta_{A^-}^{5/2}}
  \hat q\bigg(1-\frac{2\hat q^2}{5\beta_{A^-}^2}\bigg)e^{-\frac{\hat q^2}{2\beta_{A^-}^2}}\\
    \phi_{A^-}(3P,\hat q)&=\sqrt{\frac{35}{12}}\frac{1}{\pi^{3/4}}\frac{1}{\beta_{A^-}^{5/2}}
 \hat q\bigg(1-\frac{4\hat q^2}{5\beta_{A^-}^2}+\frac{4\hat q^4}{35\beta_{A^-}^4}\bigg)e^{-\frac{\hat q^2}{2\beta_{A^-}^2}}\\
   \phi_{A^-}(4P,\hat q)&=\sqrt{\frac{35}{8}}\frac{1}{\pi^{3/4}}\frac{1}{\beta_{A^-}^{5/2}}
 \hat q\bigg(1-\frac{6\hat q^2}{5\beta_{A^-}^2}+\frac{12\hat q^4}{35\beta_{A^-}^4}-\frac{8\hat q^6}{315\beta_{A^-}^6}\bigg)e^{-\frac{\hat q^2}{2\beta_{A^-}^2}},
\end{split}
\end{equation}

Now, we  treat the mass spectral equation in Eq.(\ref{s23}), which
is obtained by taking only the confinement part of the kernel, as
an unperturbed spectral equation with the unperturbed wave
functions in Eq.(\ref{wv1}). We then incorporate  the one gluon
exchange term in the interaction kernel perturbatively (as in
\cite{bhatnagar18}) and solve to first order in perturbation
theory. The complete mass spectra of ground and excited states of
heavy-light axial ($1^{+-}$) quarkonia is

\begin{equation}\label{mse01}
\frac{1}{8\beta_{A^-}^2}\bigg[M^2-(m_1+m_2)^2\bigg]+\frac{C_0\beta_{A^-}^2}{2\omega_0^2}\sqrt{1+8\hat
m_1\hat m_2A_0(N+\frac{3}{2})} +\gamma\langle V_{coul}^{A^-}\rangle
=N+\frac{3}{2},~~~N=1,3, 5,...,
\end{equation}

where $\langle V_{coul}^A\rangle$ is the expectation value of
$V_{coul}^A$ between the unperturbed states of the axial vector mesons ($1^{+-}$)
with $l=1$ and $n=0,1,2,...$, and $\gamma$ is introduced as a weighting factor to have the Coulomb term
dimensionally consistent with the harmonic term, with $\gamma$ expressed in units of $\omega_0^4/(C_0\beta_A^2)$, and it also acts as
a measure of the strength of the perturbation. The expectation value of the Coulomb term associated with the OGE term for axial $1^{+-}$
quarkonia is a single elegant expression for all states, $|nP>$, (where, $n=1,2,3,...)$,

\begin{equation}
 \langle nP\mid V^{A^-}_{coul}\mid nP\rangle =-\frac{32\pi
 \alpha_s}{9\beta_{A^-}^2}.
\end{equation}

The results of our model for mass spectrum for $1^{+-}$
$Q\overline{q}$ states along with data \cite{zyla2020}, and
other models is given in Table 1. It was observed in our previous works \cite{bhatnagar18, gebrehana19} that the mass
spectra of mesons of various $J^{PC}$ ($0^{++}, 0^{-+}$, and
$1^{--}$) is somewhat insensitive to a small range of variations
of parameter $\omega_0$, as long as $\frac{C_0}{\omega_0^2}$ is a
constant. The input parameters of our model obtained by best fit
to the spectra of ground states of scalar, pseudoscalar and vector
$Q\overline{q}$, and $Q\overline{Q}$ quarkonia are: $C_0$= 0.69,
$\omega_0$= 0.22 GeV, $\Lambda_{QCD}$= 0.250 GeV, and $A_0$=
0.01, with input quark masses $m_u$= 0.300 GeV, $m_s$= 0.430 GeV.,
$m_c$= 1.490 GeV, and $m_b$= 4.690 GeV. Using these set of input
parameters, we do the mass spectral calculations of both ground
and excited states of heavy-light axial vector ($1^{+-}$) mesons.

The numerical values of $\gamma$ multiplying $V_{coulomb}$ that gave reasonable agreement with data and other models are given in Table 1. These can at best be expressed in units of $\omega_0^4/(C_0\beta^2)$.

\begin{table}[hhhhh]
  \begin{center}
\begin{tabular}{|l|l|}
  \hline
 Mesons & $\gamma$ \\
  \hline
  $h_c(nP), \chi_{c1}(nP)$,  &                    0.085 \\
  \hline
  $c\bar{b}(nP)$           &       0.34 \\
  \hline
   $s\bar{b}(nP)$&                   0.26 \\
   \hline
   $u\bar{b}(nP)$                     & 0.26\\
   \hline
   $s\bar{c}(nP)$,    & 0.051 \\
   \hline
   $u\bar{c}(nP)$,                             & 0.051 \\
   \hline
     \end{tabular}
\caption{Values of strength of perturbation $\gamma$ in units of $GeV^2$}
\end{center}
\end{table}

We also calculated percentage contribution of coulomb term to the mass of each meson state, which are indeed small, as seen in Table 1, justifying the perturbative treatment of the coulomb term for these states. We see that for any $J^{PC}$, the contribution of coulomb term to meson mass for $b\bar{u}, b\bar{s}$, and $c\bar{b}$ mesons is larger than the corresponding contributions from $c\bar{u}, c\bar{s}$, and $c\bar{c}$ states. Also as we go to higher radial states of a given meson, the contribution of coulomb term to mass keeps decreasing from its corresponding contribution for ground states, This means that the radially excited states are loosely bound in comparison to the ground states, which is similar to the case of atoms. The mass spectra of $1^{+-}$ states are given in Table 2.

 \begin{table}[h!]
\begin{center}
\begin{tabular}{p{1.2cm} p{1.7cm} p{2cm} p{2.4cm} p{2cm} p{2cm} p{2.6cm} p{1.6cm}  }
\hline\hline
     &\footnotesize{BSE-CIA}&\% contribution of OGE &\small Expt.\cite{zyla2020}&\small BSE&\small~~ PM &\small Lattice QCD &\small~~ RQM  \\
\hline
\small $M_{h_c(1P_1)}$ &3.525&10.28\% &3.525$\pm$0.00001 &3.5244\cite{glwang} &3.581\cite{arxiv}&3.5059\cite{a1} &3.525\cite{vijaya17} \\
\small $M_{h_c(2P_1)}$&3.743&9.25\%&3.888$\pm$0.0025 &3.9358\cite{glwang} &3.934\cite{arxiv}  & &3.927\cite{vijaya17}  \\
\small $M_{h_c(3P_1)}$&3.963&8.12\%&$4.478^{+0.015}_{-0.018}$ & &  & &4.337 \cite{vijaya17} \\

\small $M_{c\bar{b}(1P_1)}$ &6.843&8.63\%& &6.8451\cite{glwang} &   & &   \\
\small $M_{c\bar{b}(2P_1)}$&7.147&8.28\%&  &7.2755\cite{glwang}  &   &   &  \\
\small $M_{c\bar{b}(3P_1)}$&7.478&7.78\%& &  & & & \\

\small $M_{s\bar{b}(1P_1)}$ &5.827&11.08\%&  &5.8364\cite{glwang}  &  &   &   \\
\small $M_{s\bar{b}(2P_1)}$&6.045&11.46\% &   &6.2803\cite{glwang}  &  &   &  \\
 \small $M_{s\bar{b}(3P_1)}$&6.299&11.27\%& &  & & &   \\

\small $M_{u\bar{b}(1P_1)}$&5.711&36.42\%& &5.7047\cite{glwang}  &  & &\\
\small $M_{u\bar{b}(2P_1)}$&5.916&11.38\% &  &6.0355\cite{glwang}   & & &\\
\small $M_{u\bar{b}(3P_1)}$& 6.159&11.35\% & & &&  &\\

\small $M_{s\bar{c}(1P_1)}$ & 2.442&14.83\% & &2.4498\cite{glwang} &  &  &  \\
\small $M_{s\bar{c}(2P_1)}$&2.605&13.47\% & &2.8304\cite{glwang}  & &    & \\
\small $M_{s\bar{c}(3P_1)}$&2.748&12.03\%&& & &        &  \\

$M_{u\bar{c}}$(\footnotesize{$1P_0$})&2.316&16.83\%&  &2.3025\cite{glwang} &  & &  \\
$M_{u\bar{c}}$(\footnotesize{$2P_0$})&2.454&15.48\%& &2.6511 \cite{glwang} && & \\
$M_{u\bar{c}}$(\footnotesize{$3P_0$})&2.601&13.88\% &  & & & \\

\hline \hline
\end{tabular}
\end{center}
\caption{Masss spectra of ground and excited states of axial
$1^{+-}$ quarkonia (in GeV) in BSE-CIA (with the percentage contribution of the OGE to meson mass) along with data and results of other models}
\end{table}
\bigskip

We now give the plots of wave functions for $1^{+-}$ states for $h_c(nP),u\bar{c}(nP), u\bar{b}(nP)$ and $c\bar{b}(nP)$, as a function of internal momentum, $|\hat{q}|$.
\begin{figure}[h!]
 \centering
 \includegraphics[width=12cm,height=6cm]{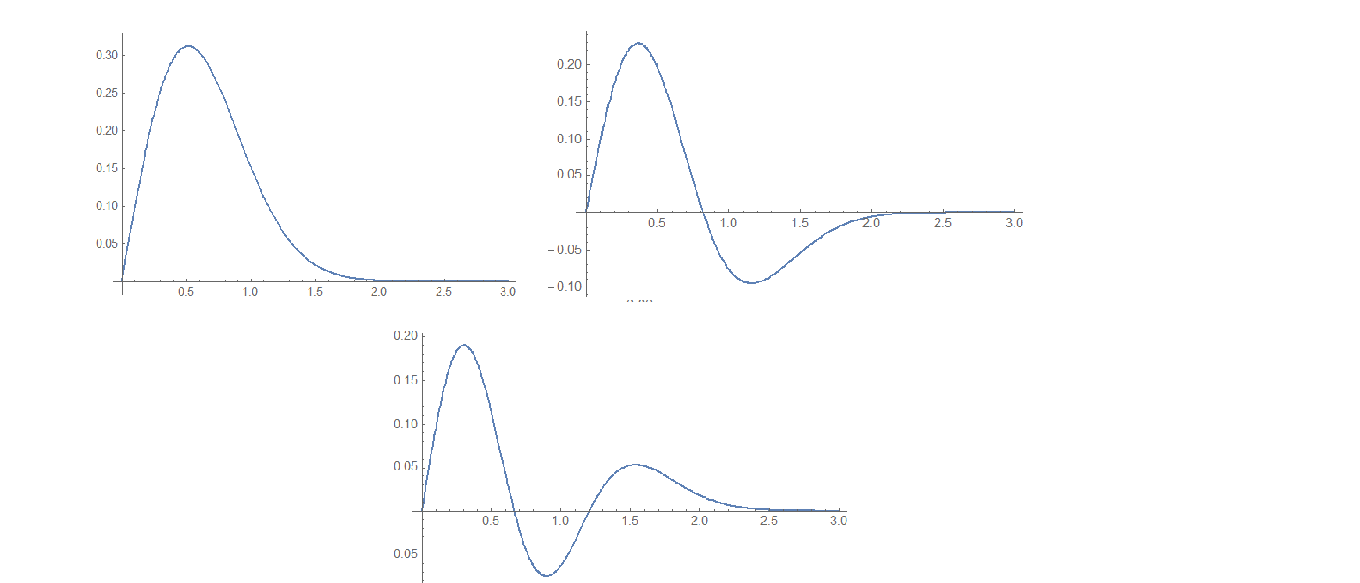}
 \caption{Plots of radial wave functions $\phi_{A^-}(\hat{q})$ for $h_c(1P), h_c(2P)$, and $h_c(3P)$ versus $|\hat{q}|$}
 \label{fig:2}
\end{figure}

\begin{figure}[h!]
 \centering
 \includegraphics[width=12cm,height=6cm]{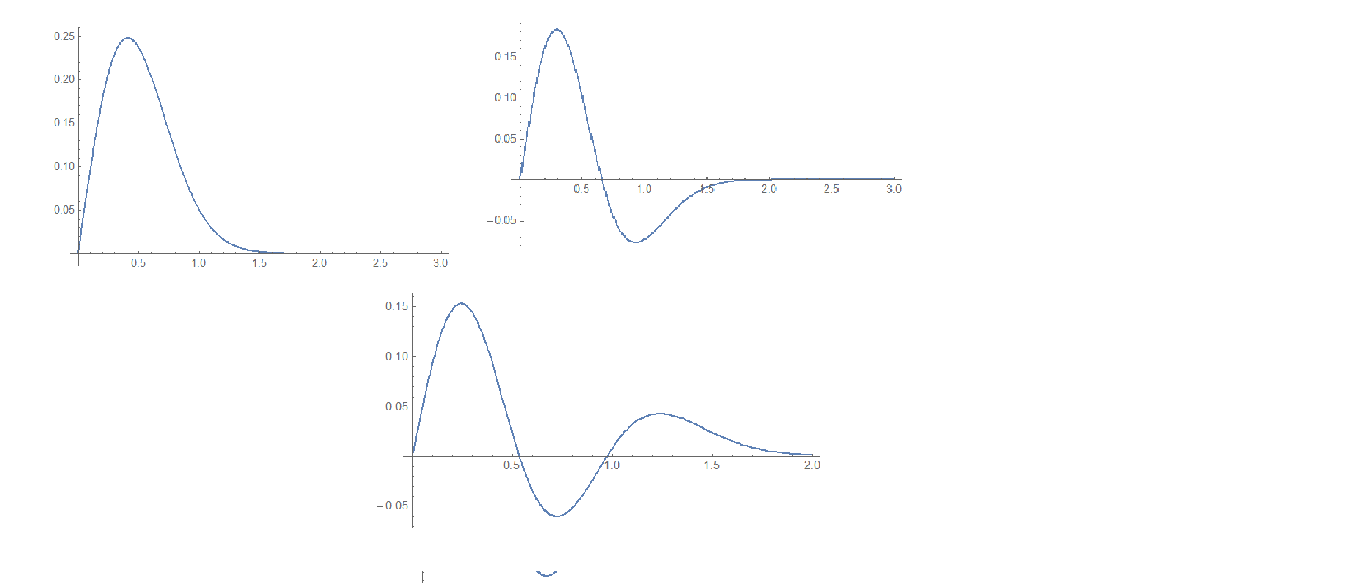}
 \caption{Plots of radial wave functions $\phi_{A^-}(\hat{q})$ for $u\bar{c}(1P), u\bar{c}(2P)$, and $u\bar{c}(3P)$ versus $|\hat{q}|$}
 \label{fig:2}
\end{figure}

\begin{figure}[h!]
 \centering
 \includegraphics[width=12cm,height=6cm]{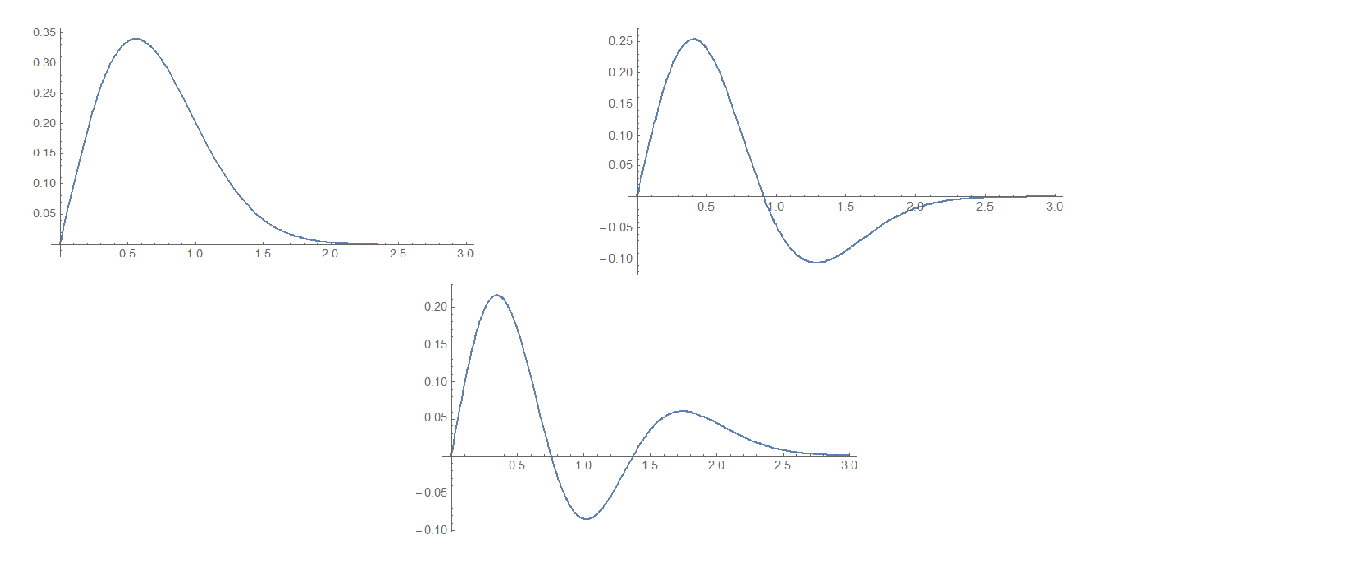}
 \caption{Plots of radial wave functions $\phi_{A^-}(\hat{q})$ for $u\bar{b}(1P), u\bar{b}(2P)$, and $u\bar{b}(3P)$ versus $|\hat{q}|$}
 \label{fig:2}
\end{figure}

\begin{figure}[h!]
 \centering
 \includegraphics[width=12cm,height=6cm]{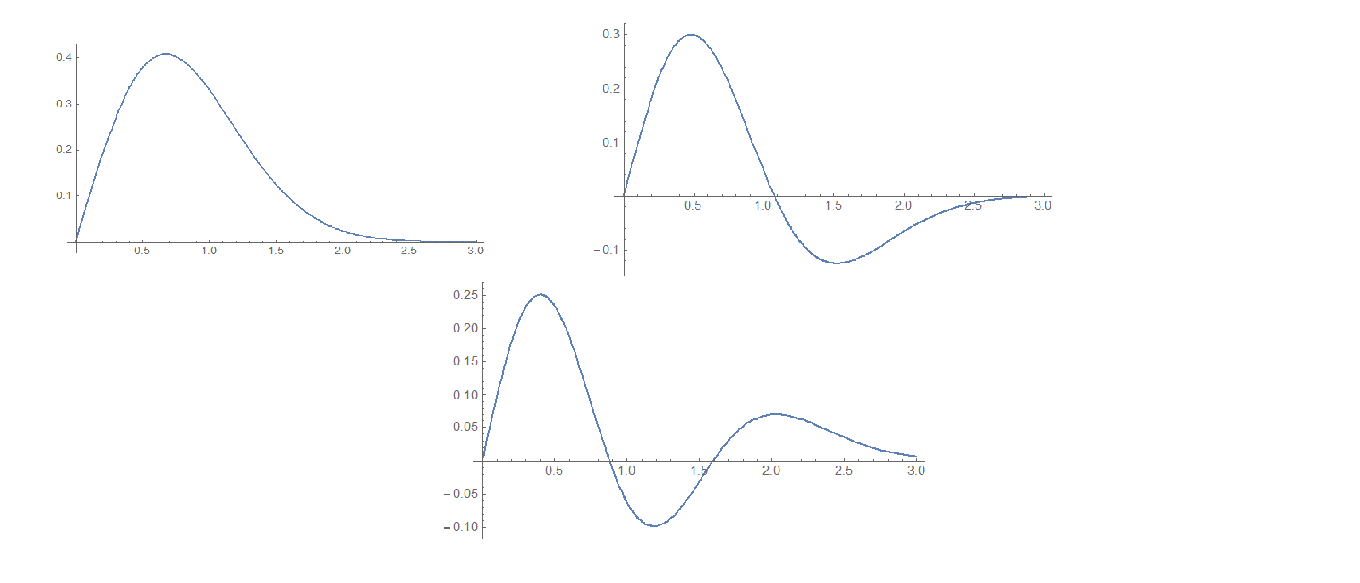}
 \caption{Plots of radial wave functions $\phi_{A^-}(\hat{q})$ for $c\bar{b}(1P), c\bar{b}(2P)$, and $c\bar{b}(3P)$ versus $|\hat{q}|$}
 \label{fig:2}
\end{figure}

We have obtained the
general expressions of 3D forms of long distance (nonperturbative) Bethe-Salpeter wave functions for $1^{+-}$ heavy-light $Q\bar{q}$ mesons. We have
given the plots of these wave functions as a function of the
internal momentum, $|\hat{q}|$ in Figs. 1–4. For $Q\bar{q}$  systems, the
wave functions show a damped oscillatory behavior. For $1^{+-}$ mesons, the amplitude of the wave function is
$0$ at $|\hat{q}|=0$ (due to the wave functions being odd), then
with increase in $\hat{q}$, it reaches a maximum, executes a
damped oscillatory behavior, and finally becomes $0$. We wish to mention that a very similar behaviour is observed for the plots of $1^{++}$ mesons, due to which we give here only the plots of $1^{+-}$ mesons.
Further, as regards all the $1^{+-}$ mesons, we see that $1P$
states have zero nodes, followed by $2P$ states with one node
and $3P$ states with two nodes.  Thus, these plots show that the 3D wave functions,
$\phi_{A^-}(nP)$ have $n-1$ nodes. This is a
general feature of all quantum mechanical systems forming a bound state. An
interesting feature of these plots is that as the mass of the
meson,$M$ increases, $\phi(\hat{q})\rightarrow 0$ at a higher value of $|\hat{q}|$.
As seen from the plots, the wave functions of heavier
mass $Q\bar{q}$ systems (such as $c\bar{b}$, $u\bar{b}$)
extend to a much shorter distance than the wave functions
of ($h_c$, $u\bar{c}$). This implies that
the heavier mesons comprising of $b$ quarks ($u\bar{b}$, $c\bar{b}$ etc.) are more tightly
bound than the comparatively lighter mesons comprising of $c$ quark ($h_c$, $u\bar{c}$ etc.).

This feature is also supported by the fact that in general,
the percentage contribution of $V_{coul}$ to meson mass, $M$, is
larger for $u\bar{b}$, than for $u\bar{c}$.

It is in this sense, the algebraic forms of 3D hadronic BS wave
functions can not only provide information about the long
range non-perturbative physics, they also tell us the shortest
distance to which they can penetrate in a hadron.

We now derive the mass spectral equations of$1^{++}$ quarkonia in the next section.

\section{Mass spectrum of $1^{++}$ quarkonia}
We start with the general form of 4D BS wave function for axial
meson ($1^{++}$) in \cite{smith,alkofer01}. Taking its dot product with $\epsilon_{\mu}$, the polarization vector of axial vector meson, and making use of the orthogonality relation, $P.\epsilon=0$, we get,

\begin{equation}
\Psi_{A+}(P,q)=\gamma_5 \not{\epsilon} [if_1(q,P)+\not P f_2(q,P)-\not{q} f_3(q,P)+i[\not{P},\not{q}]f_4(q,P)]+\gamma_5 (\epsilon.q)(f_3(q,P)+2i\not{P} f_4(q,P))
\end{equation}

Till now these amplitudes $f_{1},...,f_{4}$ in equation above
are all independent. We try to make these amplitudes dimensionless by pulling out various powers of M, and write the above expression as,

\begin{equation}
\Psi_{A+}(P,q)=\gamma_5 \not \epsilon [if_1(q,P)+\frac{\not P}{M} f_2(q,P)-\frac{\not q}{M} f_3(q,P)+i\frac{[\not q,\hat{P}]}{M^2}f_4(q,P)+\gamma_5 \frac{(\epsilon.q)}{M}(f_3(q,P)+2i\frac{\not P}{M} f_4(q,P))
\end{equation}

With the use of our power-counting rule \cite{bhatnagar06, bhatnagar11, bhatnagar14}, it can be verified that the Dirac structures associated with the
amplitudes $f_1$ and $f_2$ are $O(M)$, and are leading, and thus they would contribute the most to any axial-vector meson
calculation. Following a similar procedure as in the case of scalar mesons, we can write the Salpeter wave function in terms
of only two Dirac amplitudes: $f_1$ and $f_2$. Plugging this wave function together with the projection operators into the first two
Salpeter equations, and taking the trace of both sides and following the same steps as for the $1^{+-}$
meson case, we get the coupled integral equations in the amplitudes $f_1$ and $f_2$:

\begin{eqnarray}
&&\nonumber (M+\omega_1+\omega_2)\bigg[\frac{(m_1-m_2)}{\omega_1\omega_2}(\hat{q}.\epsilon)f_1(\hat{q})+ (\frac{1}{\omega_1}+\frac{1}{\omega_2})\hat{q}.\epsilon f_2(\hat{q})\bigg]\\&&
\nonumber =-\frac{4}{3}\int \frac{d^3\hat{q}'}{(2\pi)^3}V(\hat{q},\hat{q}')2\frac{m_1-m_2}{\omega_1\omega_2}\hat{q}'.\epsilon f_1(\hat{q}')],\\&&
\nonumber (M-\omega_1-\omega_2)\bigg[-\frac{(m_1+m_2)}{\omega_1\omega_2}(\hat{q}.\epsilon)f_1(\hat{q})+(\frac{1}{\omega_1}-\frac{1}{\omega_2})\hat{q}.\epsilon f_2(\hat{q})\bigg]\\&&
=\frac{4}{3}\int \frac{d^3\hat{q}'}{(2\pi)^3}V(\hat{q},\hat{q}')[2\frac{m_1+m_2}{\omega_1\omega_2}\hat{q}'.\epsilon f_1(\hat{q}')
\end{eqnarray}

Making use of the fact that $V(\hat{q},\hat{q}')=\overline{V}(\hat{q},\hat{q}')\delta^3(\hat{q}-\hat{q}')$ on the RHS of the two coupled equations, they convert into two algebraic equations. Following a similar procedure as in case of $1^{+-}$, the two above equations will get decoupled in amplitudes, $f_1$ and $f_2$ that are listed below.

\begin{equation}
\begin{split}
  \bigg[\frac{M^2}{4}-\frac{1}{4}(m_1+m_2)^2-\hat q^2\bigg]g_1(\hat q)&=-\frac{1}{2}(m_1+m_2)\overline{ V}_{c}(\hat q)g_1(\hat q)\\
 \bigg[\frac{M^2}{4}-\frac{1}{4}(m_1+m_2)^2-\hat q^2\bigg]g_2(\hat q)&=-\frac{1}{2}(m_1+m_2)\overline{ V}_{c}(\hat q)g_2(\hat q).
\end{split}
\end{equation}

Thus it can be checked that $f_1$, and $f_2$ satisfy the same equation for unequal mass $1^{++}$ meson, and thus, $f_1(\hat{q})\approx f_2(\hat{q})\approx \phi_{A^+}(\hat{q})$, where $\phi_{A^+}(\hat{q})$ can be shown to satisfy the mass spectral equation,

\begin{eqnarray}
&&\nonumber E_{A^+} \phi_{A^+}(\hat{q})=[-\beta_{A^+}^2\vec{\nabla}_q^2 +\hat{q}^2]\phi_{A^+}(\hat{q}),\\&&
\beta_{A^+}=(\frac{2}{3}M\omega_{q\bar{q}}^2)^{\frac{1}{4}}
\end{eqnarray}

The spectrum of $1^{++}$ is again of the $N+\frac{3}{2}$ type, with $N=2n+l$ with $n=0,1,2,...$, and $l=1$ as in the $1^{+-}$ case. The normalized odd parity energy eigen functions of Eq.(31) that are obtained by solutions of this equation are similar to $1^{+-}$, with expressions in Eq.(24), with the replacement, $ \phi_{A^-}(\hat{q})\Rightarrow  \phi_{A^-}(\hat{q})$, and $\beta_{A^-} \Rightarrow \beta_{A^+}$, with the inverse range parameter, $\beta_{A^+}$ given in Eq.(31).

The perturbative incorporation of the coulomb term into the mass spectral equation, leads to the equation,
\begin{equation}
E_{A^+} \phi_{A^+}(\hat{q})=[-\beta_{A^+}^2\vec{\nabla}_q^2 +\hat{q}^2+V_{coul}^{A^+}]\phi_{A^+}(\hat{q}),
\end{equation}
The solutions of the above spectral equation is,

\begin{equation}\label{mse01}
\frac{1}{8\beta_{A^+}^2}\bigg[M^2-(m_1+m_2)^2\bigg]+\frac{C_0\beta_{A^+}^2}{2\omega_0^2}\sqrt{1+8\hat
m_1\hat m_2A_0(N+\frac{3}{2})} +\gamma\langle V_{coul}^{A^+}\rangle
=N+\frac{3}{2},~~~N=1,3, 5,...,
\end{equation}

which leads to the mass spectra. Here, $l=1$, where $V_{coul}$ is the expectation value of $V_{coul}$ between the unperturbed states of a given quantum number, $n$ (with $l=1$) for axial mesons, with value,

\begin{equation}
<nP|V_{coul}^{A^+}|nP> =-\frac{64}{9}\frac{\pi \alpha_s}{\beta_{A^+}^2},
\end{equation}

for $1P, 2P, 3P,...$ states. The mass spectra can be calculated numerically by inverting this equation, and is given in Table 3.

 \begin{table}[h!]
\begin{center}
\begin{tabular}{p{1.2cm} p{1.7cm} p{2cm} p{2.4cm} p{2cm} p{2cm} p{2.6cm} p{1.6cm}  }
\hline\hline
     &\footnotesize{BSE-CIA}&\% contribution of OGE &\small Expt.\cite{zyla2020}&\small BSE&\small~~ PM &\small Lattice QCD &\small~~ RQM  \\
\hline
\small $M_{\chi_{c1}(1P_1)}$ &3.527&10.01\% &3.510$\pm$0.00005 &3.5244\cite{glwang} &3.581\cite{arxiv}&3.4845\cite{a1} &3.510\cite{vijaya17} \\
\small $M_{\chi_{c1}(2P_1)}$&3.811&9.23\%&3.871$\pm$0.00017 &3.9358\cite{glwang} &3.934\cite{arxiv}  & &3.872\cite{vijaya17}  \\
\small $M_{\chi_{c1}(3P_1)}$&4.147&8.41\%& 4.146$\pm$ 0.0024 & &  & &4.312\cite{vijaya17}  \\

\small $M_{c\bar{b}(1P_1)}$ &6.841&8.70\%& &6.8451\cite{glwang} &   & &   \\
\small $M_{c\bar{b}(2P_1)}$&7.171&8.54\%&  &7.2755\cite{glwang}  &   &   &  \\
\small $M_{c\bar{b}(3P_1)}$&7.553&8.20\%& &  & & & \\

\small $M_{s\bar{b}(1P_1)}$ &5.829&12.08\%&  &5.8364\cite{glwang}  &  &   &   \\
\small $M_{s\bar{b}(2P_1)}$&6.085&11.40\% &   &6.2803\cite{glwang}  &  &   &  \\
 \small $M_{s\bar{b}(3P_1)}$&6.415&11.27\%& &  & & &   \\

\small $M_{u\bar{b}(1P_1)}$&5.713&35.44\%& &5.7047\cite{glwang}  &  & &\\
\small $M_{u\bar{b}(2P_1)}$&5.961&12.56\% &  &6.0355\cite{glwang}   & & &\\
\small $M_{u\bar{b}(3P_1)}$& 6.282&12.42\% & & &&  &\\

\small $M_{s\bar{c}(1P_1)}$ & 2.460&14.23\% &2.459$\pm$0.0009 &2.4498\cite{glwang} &  &  &  \\
\small $M_{s\bar{c}(2P_1)}$&2.746&13.23\% & &2.8304\cite{glwang}  & &    & \\
\small $M_{s\bar{c}(3P_1)}$&3.042&12.00\%&& & &        &  \\

$M_{u\bar{c}}$(\footnotesize{$1P_0$})&2.313&17.83\%&  &2.3025\cite{glwang} &  & &  \\
$M_{u\bar{c}}$(\footnotesize{$2P_0$})&2.592&16.22\%& &2.6511 \cite{glwang} && & \\
$M_{u\bar{c}}$(\footnotesize{$3P_0$})&2.874&13.97\% &  & & & \\

\hline \hline
\end{tabular}
\end{center}
\caption{Masss spectra of ground and excited states of axial vector
$1^{++}$ quarkonia (in GeV) in BSE-CIA (with the percentage contribution of the OGE to meson mass) along with data and results of other models}
\end{table}
\bigskip

We now calculate the leptonic decay constants of $1^{+-}$, and $1^{++}$ mesons.
\section{Leptonic decays of $1^{+-}$ mesons}
The decay constants of $1^{+-}$ states are defined through the relation,

\begin{equation}
<0|\bar{q}_1 \gamma_{\mu}(1-\gamma_5)q_2|A^-> = (f_{A^-})M\epsilon_{\mu}.
\end{equation}

Now, this equation can be expressed as a quark-loop integral,
\begin{equation}
(f_{A^-})M\epsilon_{\mu}=\sqrt{3}\int \frac{d^3\hat{q}}{(2\pi)^3}Tr[\Psi_{A^-}(\hat{q})(1-\gamma_5)\gamma_{\mu}],
\end{equation}

with the wave function, $\Psi^{A^-}(\hat{q})$ given in Eq,(10) as,

\begin{equation}
\psi_{A^-}(\hat{q})=(N_{A^-})\gamma_5 \epsilon.\hat{q}\bigg[g_1\bigg(1- i{\not}\hat{q} \frac{(\hat{q}^2+\omega_1 \omega_2-m_1 m_2)}{\hat{q}^2 (m_1+m_2)}\bigg)+g_2\bigg(i\frac{\not{P}}{M}+\frac{(m_1\omega_2-m_2\omega_1)}{(\omega_1+\omega_2)\hat{q}^2}\frac{{\not}P{\not}\hat{q}}{M}\bigg)\bigg]\phi_{A^-}(\hat{q}).
\end{equation}

Putting the expression for $\psi^A(\hat{q})$ above, and evaluating trace over the gamma matrices, we obtain,

\begin{equation}
(f_{A^-})M\epsilon_{\mu}=\sqrt{3}\int \frac{d^3\hat{q}}{(2\pi)^3}4(\hat{q}.\epsilon)q_{\mu} \frac{(\hat{q}^2+\omega_1 \omega_2-m_1 m_2)}{\hat{q}^2 (m_1+m_2)}\phi_{A^-}(\hat{q}).
\end{equation}

Now, multiplying both sides of the above equation by the polarization vector, $\epsilon_{\mu}$, we get,

\begin{equation}
(f_{A^-})M=\sqrt{3}N_{A^-}\int \frac{d^3\hat{q}}{(2\pi)^3}4(\hat{q}.\epsilon)^2 \frac{(\hat{q}^2+\omega_1 \omega_2-m_1 m_2)}{\hat{q}^2 (m_1+m_2)}\phi_{A^-}(\hat{q}),
\end{equation}

where, the 4D BS normalizer of axial ($1^{+-}$) meson, $N_{A^-}$, can be obtained by solving the current conservation condition,

\begin{equation}
2iP_{\mu}=\int \frac{d^4 q}{(2\pi)^4} Tr\bigg[\overline{\psi}(P,q)[\frac{\partial}{\partial P_{\mu}}S_F^{-1}(p_1)]\psi(P,q)S_F^{-1}(-p_2)\bigg] +(1 \leftrightarrow 2)
\end{equation}

We make use of the fact that $S_F^{-1}(p_{1,2}=i(\pm i\not{p}_{1,2}+m_{1.2})$, where $p_{1,2}=\hat{m}_{1,2}P\pm q$. In the hadron rest frame, where $P=(\overrightarrow{0}, iM)$, and $q=(\hat{q}, i0)$, we can reduce the above equation to the 3D form,

\begin{equation}
2iP_{\mu}=\int \frac{d^3\hat{q}}{(2\pi)^3}Tr\bigg[\overline{\psi}(\hat{q})(-\hat{m}_1\gamma_{\mu})\psi(\hat{q})[-i(\hat{m}_2\not{P}+\not{q})+m_2]\bigg]+(1\leftrightarrow 2)
\end{equation}

We can express the decay constants as,

\begin{equation}
f_{A^-}=\frac{4\sqrt{3}}{M}N_{A^-}\int \frac{d^3\hat{q}}{(2\pi)^3}\frac{(\hat{q}^2+\omega_1 \omega_2-m_1 m_2)}{\hat{q}^2 (m_1+m_2)}\phi_{A^-}(\hat{q}).
\end{equation}

The values of these decay constants of $1^{+-}$ quarkonia are given in Table 4 below:

\begin{table}[htbp]
\begin{center}
\begin{tabular}{p{1.8cm} p{1.6cm} p{2.5cm} p{1.6cm} p{2.2cm} p{2.3cm} p{2.3cm} }
  \hline\hline
                  &BSE-CIA&Expt.&BSE\cite{glwang}& QCD SR1            &QCD-SR2 \\
   \hline
    $f_{h_{c}(1P)}$&0.161&  --    &0               & 0.176\cite{arxiv} & 0.490\cite{wang}          &\\
    $f_{h_{c}(2P)}$&0.094&  --    &0               &0.244\cite{arxiv}  & & \\
    $f_{h_{c}(3P)}$&0.062&  --    &0                &   & & \\

    $f_{c\bar{b}(1P)}$&0.112&    &0.050  & & &\\
    $f_{c\bar{b}(2P)}$&0.061&    & 0.049 &   &  & \\
    $f_{c\bar{b}(3P)}$&0.038 &   &       &  & & \\

    $f_{s\bar{b}(1P)}$&0.235 &   &0.076  & & &\\
    $f_{s\bar{b}(2P)}$&0.104 &   & 0.071   & &  & \\
    $f_{s\bar{b}(3P)}$&0.059&    &  &  &  & \\

     $f_{u\bar{b}(1P)}$&0.271&   & 0.076 & & &\\
    $f_{u\bar{b}(2P)}$&0.111&    & 0.070   &  &  & \\
    $f_{u\bar{b}(3P)}$&0.064&    &  &  &  &\\

    $f_{s\bar{c}(1P)}$&0.133&  &0.062 & & &\\
    $f_{s\bar{c}(2P)}$&0.094&    &0.050 &   &   & \\
    $f_{s\bar{c}(3P)}$&0.073&    &  &  &   &\\

    $f_{u\bar{c}(1P)}$&0.146&   &0.072 &  & & \\
    $f_{u\bar{c}(2P)}$&0.110 &   & 0.056 &   &   &\\
    $f_{u\bar{c}(3P)}$&0.095&    &  & &  &  \\ \hline
   \hline
   \end{tabular}
   \end{center}
   \caption{Leptonic decay constants, $f_{A^-}$ of ground
state (1P) and excited state (2P) and (3P) of heavy-light
axial vector ($1^{+-}$) mesons (in GeV.) in present calculation (BSE-CIA)
along with experimental data, and their masses in other models.}
     \end{table}

\bigskip
\section{Leptonic decays of $1^{++}$ mesons}
The decay constants of $1^{++}$ states are defined through the relation,

\begin{equation}
<0|\bar{q}_1 \gamma_{\mu}(1-\gamma_5)q_2|A^+> = (f_{A^+})M\epsilon_{\mu}.
\end{equation}

Now, this equation can be expressed as a quark-loop integral,
\begin{equation}
(f_{A^+})M\epsilon_{\mu}=\sqrt{3}\int \frac{d^3\hat{q}}{(2\pi)^3}Tr[\Psi_{A^+}(\hat{q})(1-\gamma_5)\gamma_{\mu}],
\end{equation}

where, the wave function, $\Psi^{A+}(\hat{q})$ is given as,
\begin{equation}
\Psi^{A+}(\hat{q})=N_{A^+}\gamma_5 [i\not \epsilon +\frac{\not \epsilon \not P}{M}]\phi^{A^+}(\hat{q})
\end{equation}

We put the above equation into Eq.(35) to calculate the decay constants, $f_{A^+}$, which is obtained as,

\begin{equation}
f_{A+}=\frac{4\sqrt{3}}{M}N_{A^+}\int \frac{d^3 \hat{q}}{(2\pi)^3}\phi_{A+}(\hat{q}),
\end{equation}

where, the 4D BS normalizer of axial meson ($1^{++}$), $N_{A^+}$, is again obtained by solving the current conservation conditions.

The leptonic decay constants for $1^{++}$ quarkonia are given in the Table 5 below.

\begin{table}[htbp]
\begin{center}
\begin{tabular}{p{1.8cm} p{1.6cm} p{2.5cm} p{1.6cm} p{2.2cm} p{2.3cm} p{2.3cm} }
  \hline\hline
                  &BSE-CIA&Expt.&BSE\cite{glwang}       & LFQM\cite{verma11}            &QCD-SR2 \\
   \hline
    $f_{\chi_{c1}(1P)}$&0.211& --     &0.206              & -0.105 & 0.490 \cite{wang}         &\\
    $f_{\chi_{c1}(2P)}$&0.248& --     &-0.207               &     & & \\
    $f_{\chi_{c1}(3P)}$&0.250&  --    &0.199                &    & & \\

    $f_{s\bar{b}(1P)}$&0.243 &   &0.157  &-0.166 & &\\
    $f_{s\bar{b}(2P)}$&0.173 &   & -0.156  & &  & \\
    $f_{s\bar{b}(3P)}$&0.143&    &  &  &  & \\

    $f_{s\bar{c}(1P)}$&0.159&  &0.219 &-0.159 & &\\
    $f_{s\bar{c}(2P)}$&0.137&    &-0.204 &   &   & \\
    $f_{s\bar{c}(3P)}$&0.160&    &  &  &   &\\

    $f_{u\bar{c}(1P)}$&0.166&   &0.211 &-0.177  & & \\
    $f_{u\bar{c}(2P)}$&0.129 &   &-0.143 &   &   &\\
    $f_{u\bar{c}(3P)}$&0.117&    &  & &  &  \\ \hline
   \hline
   \end{tabular}
   \end{center}
   \caption{Leptonic decay constants, $f_{A^+}$ of ground
state (1P) and excited state (2P) and (3P) of heavy-light
axial vector ($1^{++}$) mesons (in GeV.) in present calculation (BSE-CIA)
along with experimental data, and their masses in other models.}
     \end{table}

\bigskip

\section{Discussion}
We have calculated the mass spectrum and leptonic decay constants of heavy-light ($Q\bar{q}$) axial vector mesons, both $1^{+-}$, and $1^{++}$ in the framework of $4\times 4$ Bethe-Salpeter equation. We have employed a 3D reduction of the Bethe-Salpeter equation under
Covariant Instantaneous Ansatz (CIA) with an interaction kernel
consisting of both the confining and one gluon exchange terms, to
derive the algebraic forms of the 3D mass spectral equations, that are explicitly dependent on the principal quantum number, $N$. Analytical Solutions of these mass spectral equations not only leads to mass spectrum of ground and excited states of heavy-light axial vector ($1^{++}$) and ($1^{+-}$) quarkonia, but also the
eigen functions of heavy-light quarkonia in an approximate
harmonic oscillator basis. These wave functions
for heavy-light mesons so derived, are then used to calculate their
leptonic decay constants.

Exact treatment of the spin
structure $(\gamma_{\mu}\bigotimes\gamma_{\mu})$ is done in the
interaction kernel. We first derive analytically the mass spectral
equation using only the confining part of the interaction kernel
for $Q\overline{q}$ systems. Then treating this mass spectral equation as
the unperturbed equation, we introduce the One-Gluon-Exchange
(OGE) perturbatively, and obtain the mass spectra for heavy-light $1^{++}$, and $1^{+-}$ quarkonia, treating the wave
functions derived above as the unperturbed wave functions. The parameters used were fit from the mass spectrum of pseudoscalar, vector and scalar heavy-light quarkonia and are given in Section 3 in this paper.

Mass spectral calculation is an important element to study dynamics of hadrons. Further, the analytic solutions of the spectral equations also lead to
hadronic wave functions that play an important role in the calculation of various processes involving $Q\overline{Q}$, and $Q\overline{q}$ hadrons. In our calculations, the wave functions were analytically derived from the mass spectral equations in approximate harmonic oscillator basis, and were recently used to calculate the leptonic decays of heavy-light $P$ and $V$ quarkonia\cite{gebrehana19}, two photon decays of $P$ and $S$ quarkonia\cite{bhatnagar18}, and single photon radiative $M1$, and $E1$ transitions through the processes, $V\rightarrow P\gamma$, $V\rightarrow S\gamma$, and $S\rightarrow V\gamma$ \cite{bhatnagar20}. In the present work, the wave functions derived from the mass spectral equations are used to calculate the leptonic decay constants of heavy-light $1^{+-}$, and $1^{++}$ quarkonia for which there is no experimental data presently available, and can be used as a guide for experiments.

We studied the plots of hadronic
Bethe-Salpeter wave functions calculated analytically in this
work. We studied the long distance
(nonperturbative) wave functions of $1^{+-}$ mesons
as a function of the internal momentum, $|\hat{q}|$ in
Figs. 1-4.  For $1^{+-}$ mesons, the amplitude is
$0$ at $|\hat{q}|=0$ (since wave functions are odd), then
with increase in $|\hat{q}|$, it reaches a maximum. After this the amplitude shows a
damped oscillatory behavior, and finally becomes $0$. We wish to mention that very similar behaviour of plots is observed for $1^{++}$ mesons, due to which we give here only the plots of $1^{+-}$ mesons.

These plots show that the wave functions,$\phi(\hat{q})(nP)$ have $n-1$ nodes, which is a general
feature of quantum mechanical systems forming a bound state. An interesting
feature of these plots is that as the mass of the meson, $M$
increases, $\phi(\hat{q}) \rightarrow 0$ at a higher value of $|\hat{q}|$. As further seen from the plots, the wave functions of heavier mass $Q\bar{q}$ systems (such as $c\bar{b}, u\bar{b}$) extend to a
much shorter distance than the wave functions of ($h_c, u\bar{c}$), implying thereby that the heavier mesons
(comprising of $b$ quarks are more tightly bound than the comparatively lighter mesons
comprising of $c$ quark.
This feature is also supported by the fact that in general, the percentage contribution of $V_{coulomb}$ to
meson mass, $M$, is larger for $u\bar{b}$, than for $u\bar{c}$. Thus, the long distance wave functions of axial vector mesons can act as a bridge between the long distance
non-perturbative physics, and the short distance perturbative
physics. Thus the
wave functions calculated analytically by us can lead to studies
on a number of processes involving $Q\overline{Q}$, and
$Q\overline{q}$ states. These algebraic forms of wave functions are then
used to calculate the leptonic decay constants of axial vector (both $1^{+-}$, and $1^{++}$) quarkonia in this work as a test of these
wave functions.

We have first obtained the numerical values of masses for ground
and excited states of various heavy-light mesons and made
comparison of our results with experimental data and other models using the same input parameters that were used for calculation of mass spectra of scalar, vector and pseudoscalar heavy-light quarkonia, and their transitions\cite{gebrehana19,bhatnagar20}.
We then obtained the numerical values of leptonic
decay constants for these heavy-light axial vector
quarkonia with the same set of input parameters. \\

Our results of masses for ground and excited states of heavy-light axial ($1^{+-}$ and $1^{++}$) quarkonia which are in reasonable agreements with experimental data and other models. The experimental data\cite{zyla2020} for leptonic decay constants of axial vector mesons is not yet currently available, though they have been studied in some models.

We will be using the analytical forms of eigen functions for ground and excited states
of heavy-light $1^{+-}$, and $1^{++}$ quarkonia to evaluate the various transition
processes involving these quarkonia such as, $A^+\rightarrow V\gamma$, $A^-\rightarrow P\gamma$, $V\rightarrow A^+\gamma$, and $P\rightarrow A^-\gamma$ ($A^+/ A^-= 1^{++}/1^{+-}$ (axial vector), $V=1^{--}$ (vector), $P=0^{-+}$ (pseudoscalar), and $S=0^{++}$ (scalar) quarkonia) for further work.

\end{document}